\def\cm2{cm$^{-2}$}

\def\c2{C~{\sc ii}}
\def\c4{C~{\sc iv}}
\def\fe2{Fe~{\sc ii}}
\def\fe3{Fe~{\sc iii}}
\def\mg1{Mg~{\sc i}}
\def\mg2{Mg~{\sc ii}}
\def\si2{Si~{\sc ii}}
\def\si4{Si~{\sc iv}}
\def\al2{Al~{\sc ii}}
\def\al3{Al~{\sc iii}}
\def\o1{O~{\sc i}}
\def\n1{N~{\sc i}}
\def\h1{H~{\sc i}}

\def\approxlt{\mathrel{\spose{\lower 3pt\hbox{$\sim$}}
        \raise 2.0pt\hbox{$<$}}}
\def\approxgt{\mathrel{\spose{\lower 3pt\hbox{$\sim$}}
        \raise 2.0pt\hbox{$>$}}}

\documentclass[article]{has40}
\usepackage{graphicx}
\usepackage{amssymb}
\tabletypesize{\normalsize}  

\shortauthors{Stellingwerf, Nemec, \& Moskalik}
\shorttitle{RR Lyrae Blazhko Effect}

\begin{document}
\large    
\pagenumbering{arabic}
\setcounter{page}{81}

\title{The {\it Kepler} RR Lyrae SC Data Set: Period Variation and Blazhko Effect}

\author{{\noindent R. F. Stellingwerf {$^{\rm 1}$}, J.M. Nemec {$^{\rm 2}$}  and P. Moskalik {$^{\rm 3}$} \\
\\
{\it (1) Stellingwerf Consulting, 11033 Mathis Mtn Rd SE, Huntsville, AL 35803,  USA\\
(2) Department of Physics \& Astronomy, Camosun College, Victoria\\
(3) Copernicus Astronomical Centre, ul. Bartycka 18, 00-716  Warsaw, Poland} 
}
}

\email {(1) rfs@stellingwerf.com (2) nemec@camosun.ca (3) pam@camk.edu.pl}

\begin{abstract}

We study the 1.28 million short cadence {\it Kepler} data points (Q5-Q14) for RR Lyrae covering 24 cycles of the Blazhko variation, and more than one half of the 4-year modulation of the Blazhko variation. This very rich data set clearly shows variability of the 0.5667 day pulsation period (from 0.5655 to 0.5685 day) with each Blazhko cycle, as well as a variation of the Blazhko period itself (from 39.2 days at amplitude maximum to 38.4 days at amplitude minimum) during the 4 year modulation, as well as a rapid increase in Blazhko period as the new cycle commences.

\end{abstract}

\section{INTRODUCTION}

RR Lyrae may be the most familiar and most studied of all the classical variable stars. For full background and details we refer, of course, to Horace's excellent review: Smith (1995).

The {\it Kepler} data set for RR Lyrae has already yielded many new significant and surprising results, including the discovery of period doubling and period variation during the 39 day Blazhko cycle. See Kolenberg et al. (2010) for background on the data collection techniques and many of the ground-breaking new results.  Of particular interest here are any results that may pertain to the Blazhko (1907) effect, which consists of a 39-40 day modulation of the pulsation amplitude, curve shape, and pulsation phase of various features such as maximum light, ascending branch, etc.
In this presentation we concentrate on the currently available {\it Kepler} short cadence (1 minute) observations covering the releases designated Q5 through Q14. The data set Q15 was made available shortly after this meeting, and we comment briefly on its content in the last section. The Q5-Q14 data set consists of 1,277,107 flux measurements (converted to {\it Kepler Kp} magnitudes assuming the mean magnitude 7.862 given in the {\it Kepler} Input Catalog) spanning approximately 1,872 pulsation periods and roughly 24 Blazhko (39 day) periods.  The sample covers 2.7 years, which amounts to about 2/3 of the expected 4 year modulation period of the Blazhko effect.

\section{THE DATA SET}

Figure 1 shows the complete data set: {\it Kp} magnitude versus time. The time axis is Barycentric Julian Date (BJD) reduced by 2454953 days. The large, obvious variations in amplitude are the Blazhko cycles (period ~40 days). Apparently the data starts near the maximum variation phase of an approximately four year modulation (Smith et. al. 2003), and the minimum probably occurs near time point 1,000. The variations are very clear during the large amplitude range from t = 300 to 800, then become smaller and irregular with times of max variation not equally spaced and the variation somewhat uncertain. The photometry has been renormalized and scaled for consistency, and further normalization may be needed in some segments.

\begin{figure*}
\centering
\includegraphics[width=8cm,angle=-90]{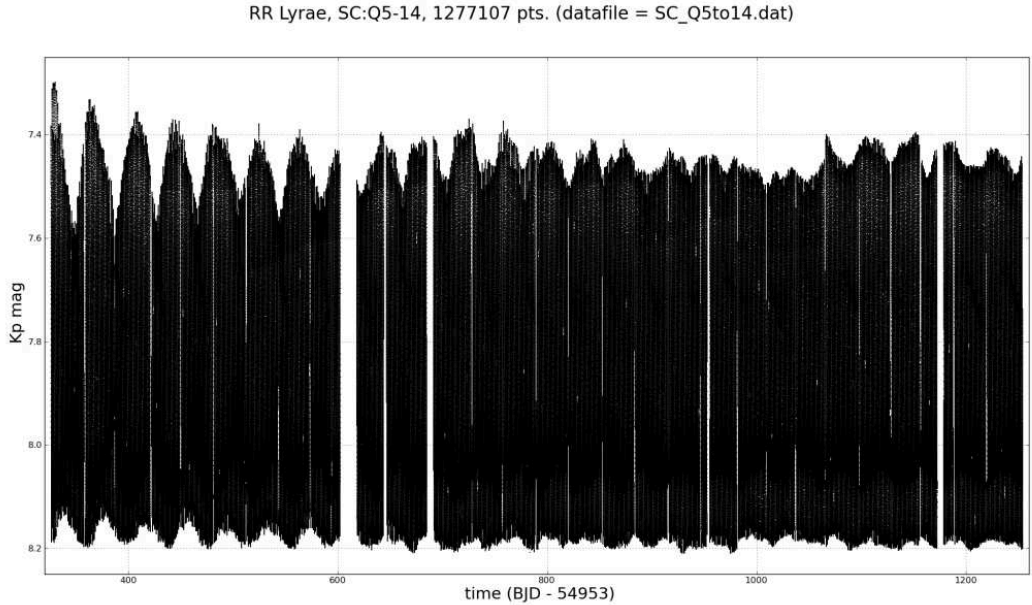}
\vskip0pt
\caption{The {\it Kepler} Q5-Q14 Short Cadence data set. }
\end{figure*}

Figure 2 shows the result of a (rolling) Fourier analysis of the data. The parameters plotted are (from bottom to top):  the amplitude and phase of the first term in the Fourier decomposition, $A_1$ and $\phi_1$; the amplitude ratio, $R_{21} = A_2/A_1$;  and the phase parameter $\phi_{31} = \phi_3 - 3 \phi_1$.  The lowest panel can be compared to the upper limit of the variation seen in Figure 1.  $\phi_1$ is the phase of this component. Note the large, regular variation of the phase – even during the last half of the plot where the amplitude is much less regular. We will show later in this paper by direct calculation that this effect is due to an actual variation of the fundamental period of oscillation during the 39 day cycle.

\begin{figure*}
\centering
\includegraphics[width=8cm,angle=-90]{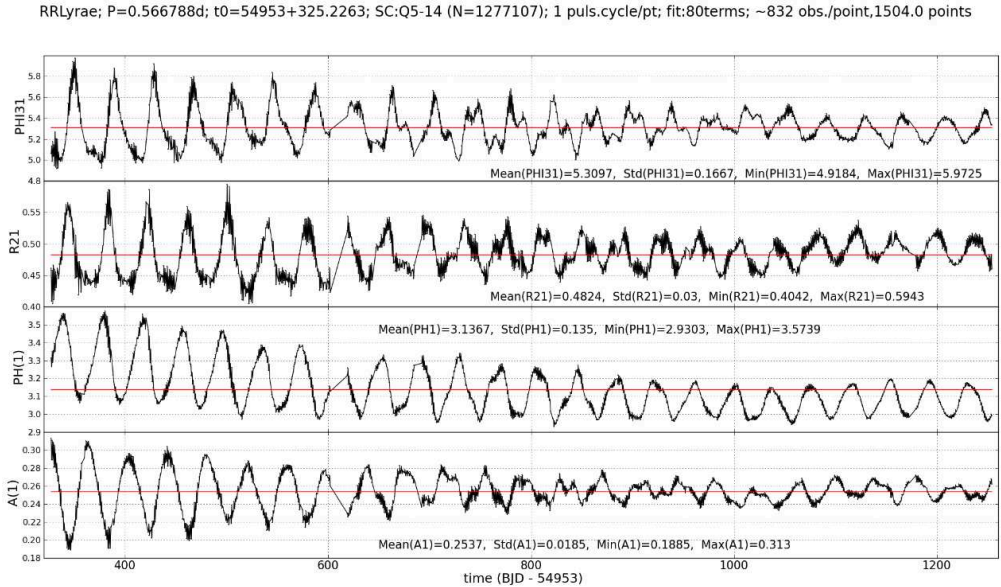}
\vskip0pt
\caption{ Time series of amplitude coefficient A1, phase coefficient ø1, amplitude ratio R21 and   ø31 phase parameter, derived from Fourier analysis of the {\it Kepler} data. See title for run parameters. }
\end{figure*}

\section{PDM ANALYSIS}

"Phase Dispersion Minimization" (PDM, Stellingwerf, 1978) was used to complement the Fourier results. Several new features are included in release "PDM2b" of the code (available at www.stellingwerf.com). They are a new "Rich Data" version of the code, developed with the assistance of Lucas Macri, and a new "Blazhko Analysis" feature, which analyzes a subset of the data, and automatically steps through the data set to determine local values and global variation of period and amplitude.

The Rich Data version is applicable to data sets with very good phase coverage. It uses 100 bins per cycle rather that the default value of 10 bins appropriate for sparse data sets. Further, this technique is usually used in conjunction with the "Spline Fit" option in which a smooth B-Spline is used to represent the mean variation, and all scatter estimates are taken relative to the mean curve, rather than to bin means as used in the original version of the technique. For a given data set, the period is chosen to minimize the scatter, and the amplitude is that of the spline fit. This results in a much smoother variation and much more precise estimates of the period and amplitude.

PDM determines the period that minimizes the scatter about a mean non-sinusoidal light variation. In this case, the technique will be applied to data segments. For each segment an accurate period and amplitude are derived. PDM does not compute a phase for each segment, so pure "phase" variation, as seen the Fourier analysis, will not be detected unless it is caused by period variation.  Since the entire light curve is used to obtain the solution, it is believed that PDM is less sensitive to changes in the shape of the light curve with amplitude that can influence the results of a Fourier or an $O - C$ analysis.

Figure 3 shows the variation versus phase for two of the segments near maximum and minimum Blazhko amplitude. Data is shown in blue, mean curve is in red. About 10 pulsation periods are represented in these analyses, so this approach averages out the period doubling effect.

\begin{figure*}
\centering
\includegraphics[scale=0.6,angle=-90]{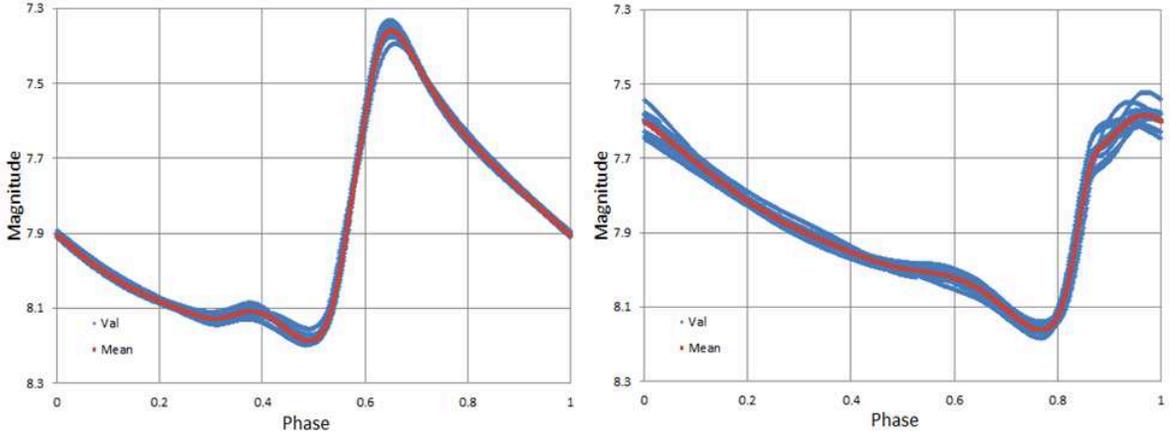}
\vskip0pt
\caption{ PDM mean curves (red)  for a segment near peak amplitude, and a segment near minimum amplitude.  The data used to derive the curves is shown in blue. }
\end{figure*}

Figure 4 shows the results of a rolling PDM analysis of the entire data set (Blazhko option) using a segment length of 10,000 points and a sequential shift of 2,500 points. Maximum, minimum, and mean values are shown for each of the 506 segments analyzed. The top and bottom curves correspond to the upper and lower variation seen in Figure 1, while the variation of the mean curve probably indicates the accuracy of the calculation. 

Figure 5 shows the amplitude versus time (magnitude at maximum light - magnitude at minimum light). The red curve is a rough mean amplitude value drawn in by eye. It is possible that some of the variations near the times 1000-1200 are due to rescaling effects. Work is underway to remedy this if needed. According to this estimate, the minimum of the Blazhko cycle occurred in the vicinity of time = 950-1050.

\begin{figure*}
\centering
\includegraphics[width=8cm,angle=-90]{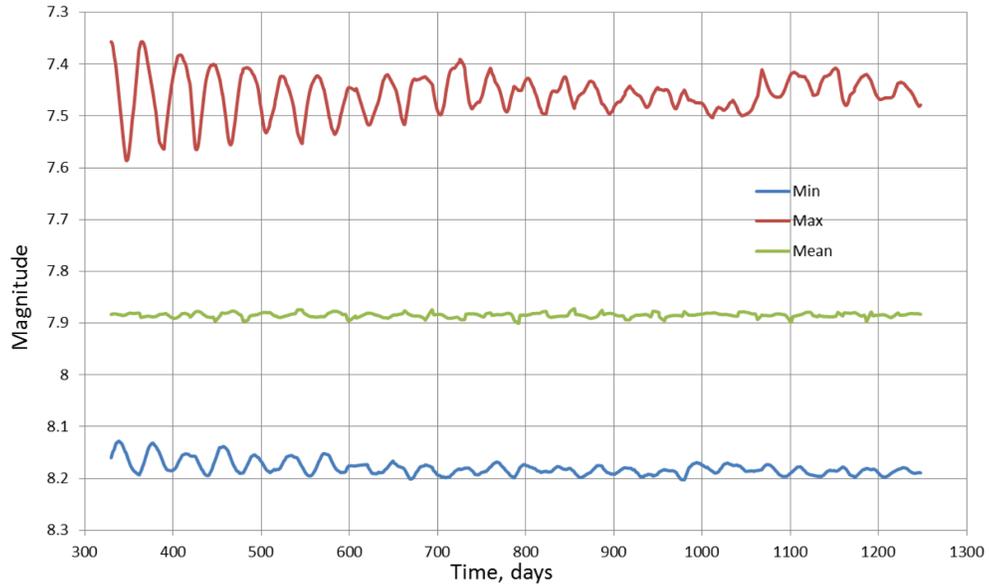}
\vskip0pt
\caption{  Magnitude maximum, minimum, and mean values versus time computed by the PDM2 analysis of the {\it Kepler} data taken in 506 segments of 10 pulsation period segments, shifted by 2.5 pulsation periods for each point plotted here. }
\end{figure*}

\begin{figure*}
\centering
\includegraphics[width=8cm,angle=-90]{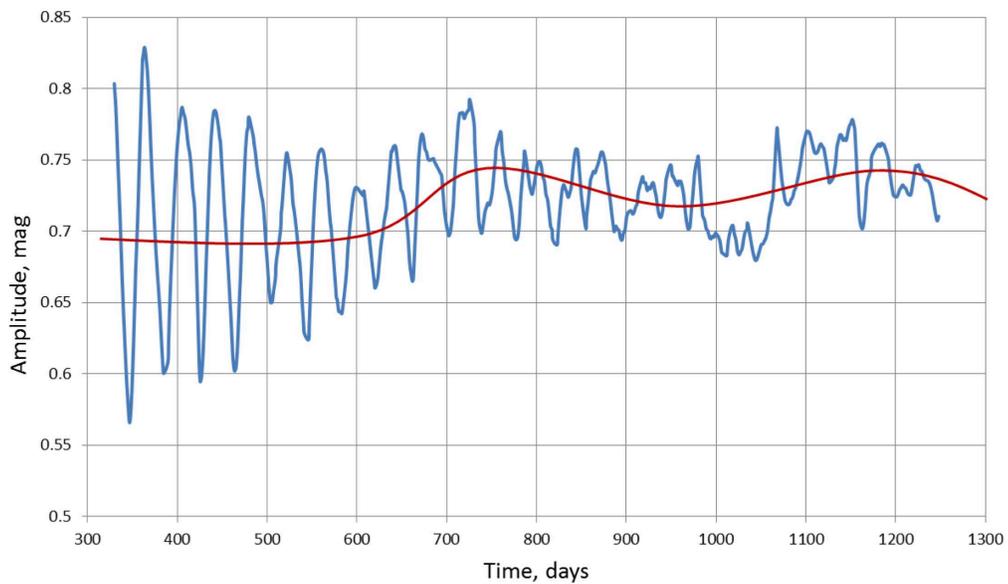}
\vskip5pt
\caption{ Amplitude ($max –- min$) versus time for the PDM analysis. }
\end{figure*}

\pagebreak

\section{PULSATION PERIOD VARIATION}

As mentioned above, another output of the PDM/Blazhko analysis is the pulsation period determined for each segment. Here we find that the period varies during the Blazhko cycle (in agreement with Kolenberg et al. (2010), who used the analytical signal method)  and that the pulsation period variation is on the whole nearly opposite to that of the amplitude variation. Figure 6 shows the period variation obtained for the entire data set. The red line is an approximate estimate of the maximum variation drawn by eye.  Figure 7 is the period versus amplitude diagram for all of the 506 segments. The negative slope is clear, as well as a rather substantial scatter. This scatter is primarily caused by counter-clockwise "loops" in the diagram, resulting from a phase shift in the period variation behind the amplitude for the large amplitude regime. We remark that other Blazhko stars show no correlation or even correlation in the opposite sense, so this result is not universal. The "high state" / "low state" implication of the shape of the red curve is noted, but will be less apparent in the analysis below. 

As an illustration of the anti-correlation of the period and amplitude variations, Figure 8 shows the two curves (rescaled and shifted for comparison) for the Q9 – Q11 segments. This is the middle portion of the data set in which the period amplitude is decreasing, covering eight Blazhko cycles. For some cycles the curves are mirror images, others are shifted, while in some cases the light variation peak appears to be truncated or inverted. This could be due to the effect of the shock wave that sometimes coincides with peak light and would tend to depress the value (presumably because of a spike in the opacity as the shock traverses the photosphere).

\begin{figure*}
\centering
\includegraphics[scale=0.6,angle=-90]{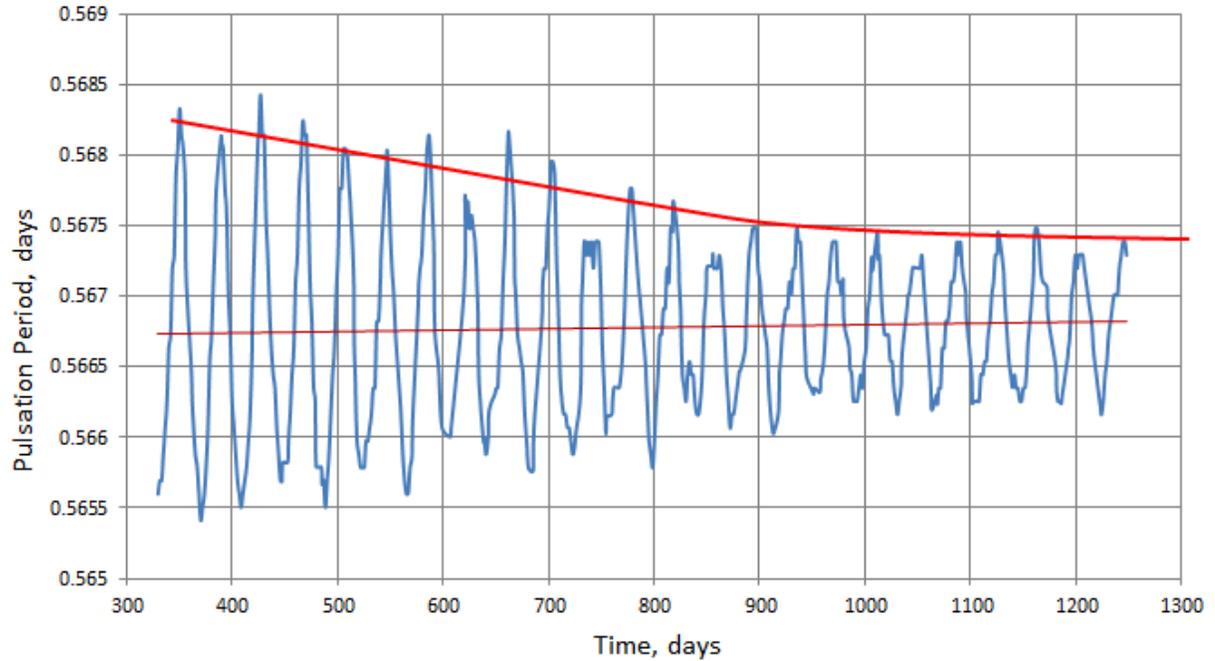}
\vskip5pt
\caption{ PDM derived pulsation period versus time. }
\end{figure*}

\begin{figure*}
\centering
\includegraphics[scale=0.55,angle=-90]{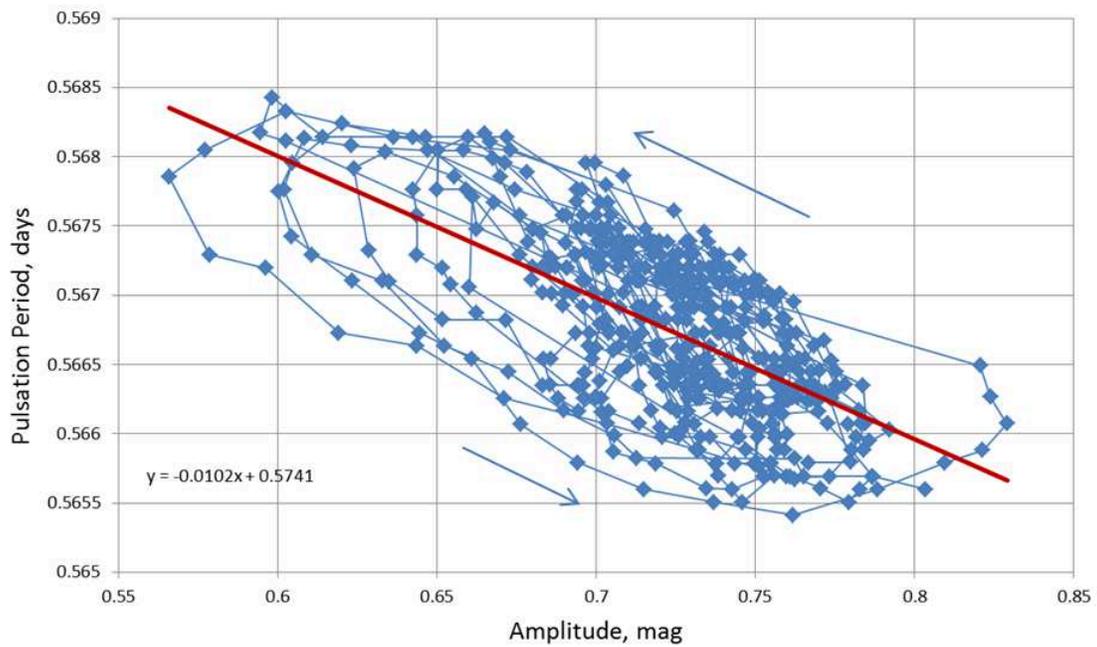}
\vskip8pt
\caption{ Pulsation period versus {\it Kp} amplitude for the PDM analysis, showing an inverse correlation and looping behavior. }
\end{figure*}

\begin{figure*}
\centering
\includegraphics[scale=0.55,angle=-90]{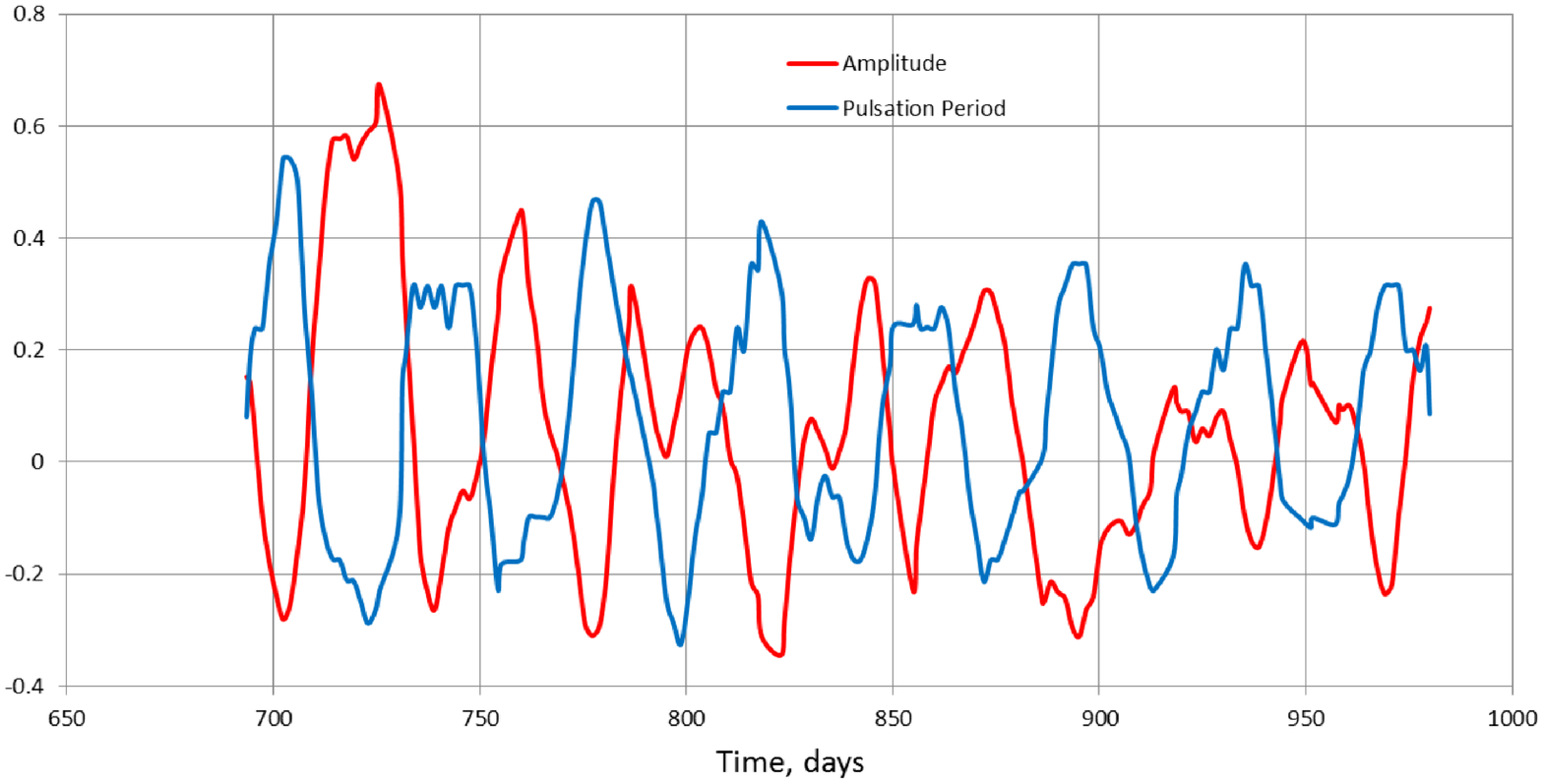}
\vskip5pt
\caption{ Rescaled pulsation period and {\it Kp} amplitude variation for SC data sets Q9-Q11. }
\end{figure*}

The period variation at the beginning of the data set is 0.53\%. This is huge when compared to the accuracy of the "known" period of RR Lyrae, currently computed to be 0.566788 d  (Table 1 of Nemec et al. 2013).  In fact, to be accurate, the period should be given as 0.566788 $\pm$ 0.0015 d, where the latter term does not represent uncertainty but known variation every 40 days. The mean period found here is close to 0.5666 d, which is very close to the period 0.5668 d obtained by Shapley (1916).  It thus appears that the mean period has varied very little, if any, and the large variation during the Blazhko cycle may account for occasional "variations" in the period found from time to time over the past century.

The cause of this observed period variation is not known. One candidate is the effect of period dependence on amplitude (Stellingwerf 1975, Tables 1-4). This effect can be positive or negative, depending on the model.  Whatever the cause we note that the period continues to vary clearly and regularly during the last half of the data set when the light amplitude is small and irregular. This suggests that the Blazhko effect is rooted deep in the stellar envelope and is caused by some mechanism capable of affecting the pulsation period, while the light amplitude is susceptible to surface irregularities of some sort that tend to obscure the amplitude variation during the minimum of the 4-year modulation.

\section{BLAZHKO PERIOD VARIATION}

Since the period variation is so well determined, we can use it to study the longer-term Blazhko variation. Figures 9 and 10 show the results of a PDM (Blazhko) analysis of the 506 point period data set taken 150 points at a time (about seven Blazhko cycles) and shifted by 10 points for each analysis. Using the segmentation option on such a short series of points is less optimal than the full analysis discussed above, and is susceptible to "edge effects" as individual points at the ends of each segment enter or drop from the analysis. Using a small shift value increases the number of points in the resulting analysis and clearly shows the size of the numerical scatter. 
Figure 9 shows the resulting minimum, mean, and maximum values for the period variation obtained in this manner -- compare with Figure 6. Figure 10 shows the derived values for the Blazhko period for each segment of the period data. Here the blue dots are the PDM results and the green line is a polynomial least squares fit to this data. Clearly, the Blazhko period has decreased from about 39.2 d to 38.4 d over the period analyzed, and then shows some tendency to increase at the end. The overall conclusion is that the Blazhko period shows a clear decrease over the 2 year period of decreasing amplitude of about 2\%. Note that the dependence on amplitude is opposite to that of the pulsation period found above. This is a large variation, and, again, may contribute to the irregular variation of the historical Blazhko period reported in the literature (e.g. Kolenberg 2008).

This result was unexpected, and some validation was desired. To this effect the data set was divided into 12 overlapping segments (about 4 Blazhko periods each), each of which was analyzed using an optimum fit of 34 Fourier components involving the pulsation period, the Blazhko period and various interaction periods. The Blazhko periods so obtained are plotted as red circles in Figure 10. The PDM and Fourier analyses show good agreement of both the Blazhko period values and the degree of scatter.   Since the Fourier method is a direct analysis, as opposed to the 2-step PDM approach, the two ends of the data sequence are better represented. Of particular interest is the last Fourier point. This point suggests that the Blazhko period is rapidly increasing while the amplitude remains low.

\begin{figure*}
\centering
\includegraphics[width=8cm,angle=-90]{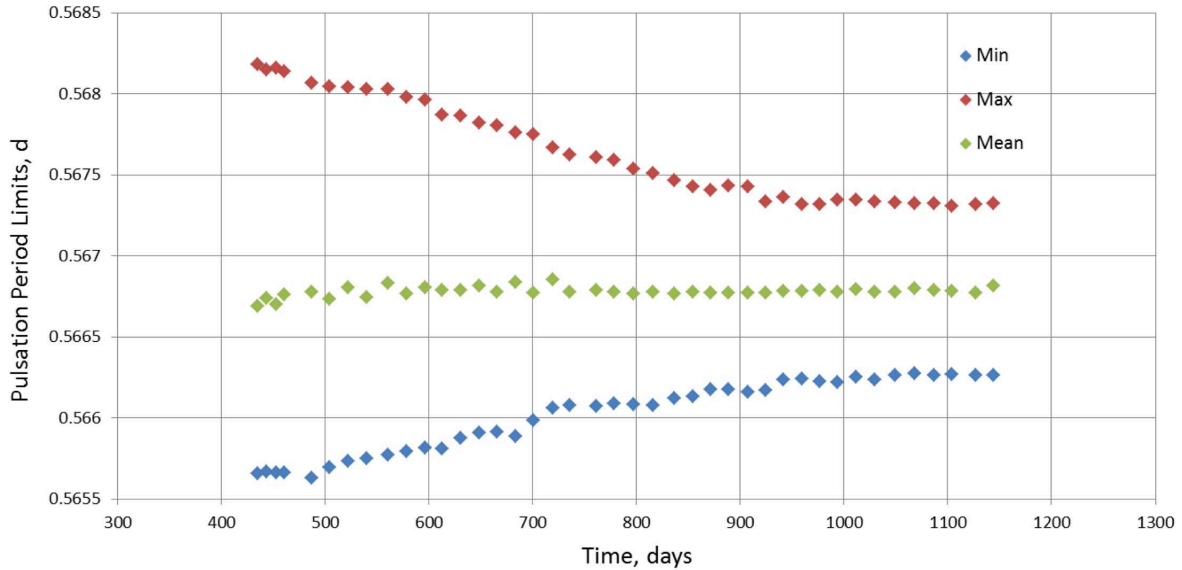}
\vskip0pt
\caption{ PDM analysis of the period variation (150 period points in a segment, shifted by 10 points each plotted point).  Limits of the pulsation period variation. }
\end{figure*}

\begin{figure*}
\centering
\includegraphics[width=8cm,angle=-90]{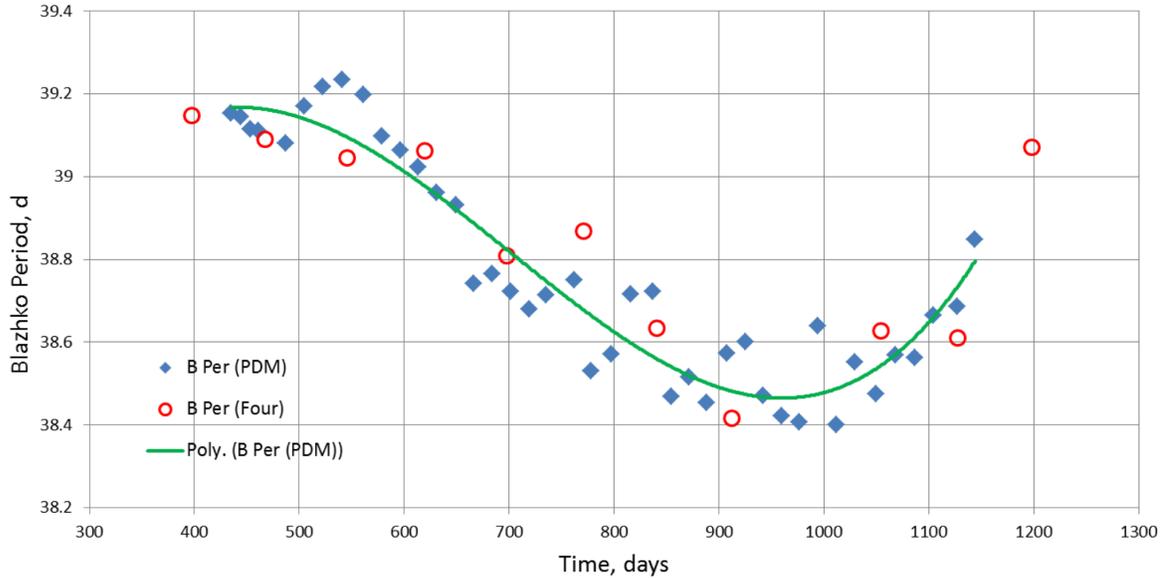}
\vskip10pt
\caption{ PDM analysis of the period variation (150 period points in a segment, shifted by 10 points each plotted point. Blue points = derived Blazhko period variation, green line = polynomial fit to blue points, red circles = results of a Fourier analysis using 12 overlapping data segments. }
\end{figure*}

\section{Q15}
After the presentation of these results, but before publication of this memo, the Q15 {\it Kepler} data set has become available. Publication of the Q15 data and results will appear separately, but we can mention here that the Q15 data set strongly supports the upward trend of the Blazhko period seen in Figure 10, and extends the upward variation to a possible Blazhko period of about 39.5 d at time 1300.

\section{SPECULATION AND PREDICTION}

Although the main conclusion of this analysis is that the Blazhko period has decreased by 2\% as the Blazhko amplitude has decreased, the sharp upward trend at the end of the data set suggests that this is not the entire story. If the upward trend continues, the Blazhko period may increase to the vicinity of 40 d at the beginning of the new cycle, thus attaining levels not seen since mid-20th century. Further, if this sort of variation occurs every Blazhko cycle, then the accepted historical trend of decreasing Blazhko period may have to be revisited.

Variation of the Blazhko period is generally taken to be evidence against its interpretation as a rotation period. We note, however, that there is one rotation scenario that is consistent with this observed variation – {\bf differential rotation effects.} If the Blazhko effect resembles the solar cycle, it may be caused in some way by a mechanism appearing at high stellar latitudes, as in the sunspot cycle of the sun, and move to lower latitudes as the cycle progresses. If the star possesses a differential rotation field that causes longer periods at higher latitudes, as in the sun, then a variation in cycle period can be expected. In the sun this causes up to a 20\% variation in rotation period as measured by sunspots. For RR Lyrae, 2\% will suffice. This resembles the mechanism suggested by Detre and Szeidl, 1973, although the 10 d phase shift seen in their O-C plots has not been detected in the RR Lyrae data to date.

At this stage this is simply an idea, and far from proven. How a magnetic disturbance affects the pulsation amplitude, and whether the perturbation is caused by star spots, an oblique field, or some other effect is not clear. If this idea is true, however, we can make the following {\bf prediction: the Blazhko cycle period should increase rapidly as the new cycle begins with a high latitude disturbance, then decrease as the amplitude increases for the first half of the cycle, followed by the observed decrease in cycle period as the amplitude decreases in the second half of the cycle.}

\section{CONCLUSIONS}

The main conclusions from our analysis of the Q5-Q14 short cadence {\it Kepler} photometry are:

1. The {\em pulsation period} varies about 0.5\% every 39 days – the period is shorter at peak amplitude, larger at small amplitude.
2. This period variation persists during the minimum of the Blazhko cycle, even when the light variation is small and irregular.
3. The mean pulsation period is close to the value measured in 1916.
4. The Blazhko amplitude variation has decreased from a maximum of about 40\% in light amplitude variation to a minimum of 0-10 \% light amplitude.
5. The {\em Blazhko period} has decreased by about 2\% from a value of 39.2 d at the peak of the cycle to a minimum of about 38.4 d at cycle minimum.
6. The Blazhko period appears to be increasing rapidly at the beginning of the new cycle.
7. The {\em modulation period} is consistent with a 4 year meta-cycle length.

In final conclusion, we extend our warmest congratulations to Horace Smith on this important occasion, and wish him all the best in his future endeavors.

\end{document}